%% file: Edge_server_deployment_scheme_of_Blockchain_in_IoVs.tex
\documentclass[journal,10pt]{IEEEtran}

\ifCLASSINFOpdf
\else
\fi

\usepackage{stfloats}
\usepackage[fleqn]{amsmath}

\usepackage[noend]{algpseudocode}

\usepackage{algorithmicx,algorithm} 

\usepackage{graphicx,epsfig,color}

\usepackage{enumerate}
\input{mac90}

\squeeze

\newcommand{\complexity}{O({1\over
\epsilon\gamma^2}({\ln{{D\over m_0}}})(\ln\ln({D\over
m_0})+\ln{1\over \epsilon})\cdot T(M, n_D, {2D \over m_0}))}

\newcommand{\complexitytwo}{O({1\over
\epsilon\gamma^2}({\ln{{D\over m_0}}})(\ln\ln({D\over
m_0})+\ln{1\over \epsilon})\cdot n^2\log n)}

\newcommand{\format}{png}

\begin{document}
%
\title{Edge server deployment scheme of blockchain in IoVs}
%
%
%

\author {Liya Xu, Mingzhu Ge, Weili Wu, ~\IEEEmembership{Member,~IEEE}\\

\thanks{ Corresponding\quad author: Mingzhu\quad Ge}
\thanks{Liya Xu, School of Information Science and Technology, Jiujiang University, jiujiang, 332005, China, e-mail: xuliya603@whu.edu.cn}
\thanks{Mingzhu Ge, Department of Information Technology Center, Jiujiang University, Jiujiang, 332005, China, e-mail: mingzhug1989@gmail.com}
\thanks{Weili Wu, Department of Computer Science, University of Texas at Dallas, Richardson, TX 75080, USA, e-mail: weiliwu@utdallas.edu}
}

{}
\maketitle

\begin{abstract}
With the development of intelligent vehicles, security and reliability communication between vehicles has become a key problem to be solved in Internet of vehicles(IoVs). Blockchain is considered as a feasible solution due to its advantages of decentralization, unforgeability and collective maintenance. However, the computing power of nodes in IoVs is limited, while the consensus mechanism of blockchain requires that the miners in the system have strong computing power for mining calculation. It consequently cannot satisfy the requirements, which is the challenges for the application of blockchain in IoVs. In fact, the application of blockchain in IoVs can be implemented by employing edge computing. The key entity of edge computing is the edge servers(ESs). Roadside nodes(RSUs) can be deployed as ESs of edge computing in IoVs. We have studied the ES deployment scheme for covering more vehicle nodes in IoVs, and propose a randomized algorithm to calculate approximation solutions. Finally, we simulated the performance of the proposed scheme and compared it with other deployment schemes.
\end{abstract}


%
\IEEEpeerreviewmaketitle

\section{Introduction}
%
%
%
%
\IEEEPARstart{w}{ith} the continuous improvement of the intelligent level of vehicles, the development of the Internet of vehicles is accelerating. The interaction and sharing of data in the Internet of vehicles has become a hot research topic. The data of vehicle interaction includes road information, data generated by vehicles, data transmitted by other nodes, etc. How to ensure the safe transmission of this information is a challenge for the development of the Internet of vehicles and the Internet of things\cite{novo2018blockchain}. Blockchain is considered to be a good solution for information security transmission due to its advantages of unforgeability, traceability, collective maintenance, etc. All devices maintain a blockchain, information is transparent, and information can be exchanged safely between different devices\cite{li2019computing}\cite{chen2019cooperative}. It is challenge to apply blockchain in the Internet of vehicles to ensure the safe transmission of data \cite{yang2018blockchain}.

%

\par The key process of blockchain technology is a computing process called "mining". It needs strong computing power to solve the proof-of-work puzzles, which takes a long time. This seriously restricts the application of blockchain in mobile Internet such as the Internet of vehicles\cite{kang2019toward}. Because the computing power of a single mobile device is often unable to satisfy this magnitude of difficult computing. As an extension of cloud computing, edge computing has gradually attracted people's attention\cite{satyanarayanan2017emergence}\cite{ren2019collaborative}. Edge computing provides an open platform integrating network computing and network storage for the real-time nearest service of user. It is initiated on the network edge, which can produce faster response to network service and satisfy real-time requirements. Edge computing can provide computing power, data storage, application services, etc\cite{luo2020edge}.

\par Therefore, it is an inevitable trend to adopt edge computing to implement the application of blockchain in mobile Internet such as the Internet of vehicles\cite{li2020energy}\cite{zhao2019computation}. The key entity of edge computing is the edge servers(ESs), which can be considered as a miner in the blockchain of the Internet of vehicles. Edge servers are connected with each other, and each edge server has large computing power and storage capacity, which can deal with data in real time. They act as the blockchain manager to perform the creation and verification of the block data. The edge servers compete with each other for the right to package data by mining, and the winner adds his own block data to the blockchain\cite{sun2019joint}. In IoVs, It can organize a large number of vehicles and other mobile devices to share the computing tasks for mining, which can greatly improve the computing power of the miner. Roadside units(RSUs) has more stable network topology, more reliable communication channels, and more powerful computing and storage capabilities than vehicle nodes. Therefore, RSU can be considered as an edge server in the environment of Internet of vehicles. It is as the miner competing for mining tasks. In addition, it collects the information uploaded by vehicles to assist information transmission. Then, the optimal deployment scheme of the edge server to cover as many vehicle nodes as possible to satisfy the coverage and connectivity of IoVs is the problem to be solved in this paper.

The main contributions of this work are summarized as follows.
\begin{itemize}
\item 
This paper introduces the important role of blockchain technology in information security transmission, as well as the challenges of blockchain application in the IoVs. In addition, we analyze the feasibility of employing edge computing to realize the application of blockchain in the IoVs.
\item 
We consider the roadside unit as the edge server, and propose a random deployment algorithm of the edge server for the blockchain in IoVs to satisfy the coverage of the edge server to the vehicle nodes
\item 
A simulation algorithm that contains a rigorous analysis is developed for performance evaluation. In addition, we simulated our scheme and compared it with another scheme.
\end{itemize}

\par The rest of this paper is organized as follows. Related work is briefly introduced in Section~\ref{related-work-section}. The randomized algorithms is presented in Section~\ref{algorithm-section}. We develop a
simulation algorithm that contains a rigorous analysis in
Section~\ref{simulation-algorithm-section}. The performance evaluation is given in Section ~\ref{simlation-section}, followed by conclusions.

\section{Related work}\label{related-work-section}
The integration of edge computing and blockchain is an inevitable way to expand the application of blockchain in mobile Internets. The architecture of edge computing or edge server deployment scheme is one of the important components to implement the edge computing. There have been several studies on the architecture for edge computing in IoT. 

\par Zheng et al\cite{zheng2018microthingschain} proposed a blockchain based distributed architecture named MicrothingsChain. The edge servers are designed to have powerful computing and storage capabilities, which can implement the interaction of Internet of things data and distributed storage of massive data. Due to the distributed storage and non-tampering characteristics of blockchain, data security and cross domain access of users can be guaranteed. Damianou et al\cite{damianou2019architecture}analyzed the challenges for the design of mobile blockchain edge computing architecture, as well as the differences with the traditional blockchain architecture, and proposed a new architecture that can reduce the storage capacity requirements of IOT devices and improve the overall performance. Sharma et al \cite{sharma2017software}designed a secure distributed fog node architecture based on blockchain technology. Fog nodes are considered as the edge servers in edge computing. They are deployed on the edge of the IoTs to respond to the access requirements of IoT devices in real time. It provides low-cost and secure computing services for devices in Internet of things. 

\par Zhang, Li and Cui\cite{zhang2018security} proposed a mobile edge computing based architecture in VANET by employing the security of blockchain. The architecture is composed of three layers. The three layers from bottom to top is perception layer, edge computing layer and service layer. The bottom layer collects and uploads data to the middle layer. The middle layer acted by edge computing layer that processes and stores data to provide data services for the service layer(the top layer). The service layer receives data and employs blockchain technology to guarantee security. Zhu, Huang and Zhou\cite{zhu2018edgechain} propose edge architecture named edgechain in blockchain based on minimizing the deployment cost of mobile edge server. They employ random programming scheme to study the deployment cost of edge server, so as to provide users with edge computing services. 

\par The Roadside Units are considered as the edge servers of edge computing in mobile blockchain under the environment of IoVs. Therefore, the deployment of edge servers is similar with the RSUs deployment in IoVs. Many researchers have studied the deployment scheme of RSUs in IoVs. Peng and Qin\cite{li2015delay} have proved that the problem of RSUs deployment is NP-hard. They deployed the RSUs by a greedy idea and two-phase scheme to obtain an approximate optimal solution. Younghwa and Jaehoon\cite{jo2016rpa} deployed the RSUs in intersection. A GSC Algorithm was developed to choose the intersection of roads. So, the scheme of the RSUs placement is the selection of intersection in roads. the scheme proposed in \cite{hwang2019efficient} is similar with them. However, it restricts the location of RSUs that can be deployed. In \cite{saravanan2019augmented}, The author integrates Powell's mathematical model with the bionic algorithm krill herd and proposes a novel RSU deployment algorithm, which adapts to the scene of sparse nodes in the Internet of vehicles and satisfies the collection of information. Zhenyu et al\cite{wang2017centrality} proposed a centrality-based RSUs placement scheme. They formulated the problem of RSUs placement as the problem of linear programming. The purpose was to maximize the number of location choose for RSUs placement under the cost of placement given.

\section{Approximation Scheme}\label{algorithm-section}

 In this paper, a randomized approximation algorithm is presented about the edge servesr(ESs) deployment for blockchain of IoVs in this section.

\subsection{Network Model}

\par The ESs are deployed on the side of the road. Vehicles node are distributed randomly on a highway and the speed of vehicles is within the given range. There are two connection ways that each vehicle communicates with ESs: 1)access directly to ESs; 2)access to ESs by multi-hop relaying. Vehicles forward information to the ES in the same direction of vehicle moving rather than the opposite direction of vehicle moving. We assume that all vehicle nodes and ESs have the same transmission scope $m_0$. It is similar with the network model in \cite{liya2013randomized}.

\subsection{Problem Description}

\par Due to the high dynamic topological structure in IoVs, the frequent breakage of link disrupts the transmission of information. The deployed ESs should be able to receive the information uploaded by vehicles and assist information transmission. For simplicity, the distance of deploying ESs is equal in this paper. We need get the optimal deploying distance of ESs, which can transmit information in IoVs with the connectivity probability $p_0$ within the time $t_0$.

\par Assume that a message can be transmitted to a vehicle of distance at
most $m_0$, the speed on the road is $v_0$, the average number of
vehicles is $d_0$ per kilometer. The Chernoff bound\cite{motwani2010randomized} will be adopted to analyze this algorithm.


{\bf Proposition 1\cite{motwani2010randomized}. } Define $X_1,\cdots, X_n$ to be independent random variables, and the value of each variable is 1 or 0. $X_i$ takes $1$ with probability $p_i$. Let $X=\sum_{i=1}^n
X_i$, and $\mu=E[X]$. Then for any $\delta>0$,
\begin{enumerate}
\item
$\Pr(X<(1-\delta)\mu)<e^{-{1\over 2}\mu\delta^2}$,
\item
$\Pr(X>(1+\delta)\mu)<\left[{e^{\delta}\over
(1+\delta)^{(1+\delta)}}\right]^{\mu}$.
\end{enumerate}

{\bf Proposition 2\cite{li2002closest}. } Define $X_1,\cdots, X_n$ to be independent random variables, and the value of each variable is 1 or 0, and $X=\sum_{i=1}^n X_i$.
\begin{enumerate}
\item
 If $P_i(X_i=1) \leq p$, then for any $\epsilon>0$, $\Pr(X>pn+\epsilon
n)<e^{-{1\over 3}n\epsilon^2}$.
\item
 If  $P_i(X_i=1) \geq p$, then for any
$\epsilon>0$, $\Pr(X<pn-\epsilon n)<e^{-{1\over 2}n\epsilon^2}$.
\end{enumerate}


{\bf Definition 1.} Assume that each ES has a unique identification
number $x$.

\begin{itemize}
\item
 A {\it connection topology} of a set of ESs is defined
by a function $g:N\rightarrow N$ such that for two ESs with
identification numbers $x$ and $y$, they are connected if and only
if $g(x)=g(y)$.

\item
If all ESs are connected with wires, then they can use the function
$g_c(x)=1$ for each ES with identification $x$.

\item
If all ESs are isolated without wire connection, then they can use
the function $g_u(x)=x$ for each ES with identification $x$.
\end{itemize}

{\bf Definition 2.} The $M$ is a set that contains the parameter of highway traffic property such as node transmission range $m_0$, vehicle speeds range $[v_1, v_2]$, the average number $b$ of nodes per unit and so on.

{\bf Definition 3.} Let $M$ be a set of parameters. Parameters $d>0, q\in [0,1], D>0$. Let $g(.)$ be a
connection topology.  Let the accident site send out a message to be
relayed over a IoVs.  Define the following random events.
Let R$_g(d,q, M,D)$ be a random event within interval ES distance $d$, and has function $g(.)$ for its ES connection
topology. It returns $1$ if one packet can be transmitted to $qn$ vehicles. The $n$ indicates the number of vehicle nodes that exist in the area of distance $D$ to the given site.




{\bf Definition 4.} Let $p,q\in [0,1]$. Let $n_D$ be the number of
vehicle nodes with the distance $D$ to the source that sends a message. Let
$M$ be a parameter set for the highway. Let
$g:N\rightarrow N$ be a connection topology. Let $f_g(p,q, M,D)$ be the largest distance $d_{\max}$ such
that for each $d\in [0,d_{\max})$, if ESs are arranged with
distance $d$ between two consecutive ESs via connection
topology $g(.)$ on a highway, it  guarantees that with at least
probability $p$, at least $qn_D$ vehicles within distance $D$
receive the message.



\subsection{Randomized Algorithm}

In this section, we introduce a random algorithm to calculate an
approximate distance for deploying edge servers. Its correctness and
computational complexity are proved.


 We discuss an algorithm framework that is suitable for both
connected ESs via some wired network and unconnected ESs network.
We propose an approximation scheme for edge servers placement and
configuration in highway scenarios. The algorithm iteratively calculates an approximate deployment distance for ESs by approaching
the optimal distance from the initial distance $m_0$. The $m_0$ is the maximum distance of node wireless transmission. If the IoVs cannot meet the conditions, then increase sequentially the distance to $m_0(1+\theta),m_0(1+\theta)^2,...,m_0(1+\theta)^i, ...$
until the IoVs meets the conditions at distance
$m_0(1+\theta)^{i+1}$, where $\epsilon$ is a precision parameter adopted to regulate the approximation to the optimal deployement distance for ESs. Then $m_0(1+\theta)^{i}$ is the approximate optimal deployment distance for ESs. For each distance
$d_i=m_0(1+\theta)^i$, We sample the sufficient number $t$ of random events, which exist in the area of distance $D$ to the given site. The events that meet the condition of information transmission on the highway will be counted. We make that with probability close to $p$, at least
$qn_D$ vehicles can receive the message ($n_D$ indicates the number
of vehicles that exist in the area of distance $D$ to the given site), the
Chernoff bound is adopted to ensure the probabilistic approximation
to $p$. The algorithm returns a distance $d$ in the range
$[{f_g(p+\gamma,q,M,D)\over 1+\epsilon},f_g(p-\gamma,q, M,D)]$ as an
approximation to $f_g(p,q, M,D)$.

{\bf Definition 5.} Let $M$ be a parameter for the highway, and let $g(.)$
be a connection topology. They satisfy {\it monotonic condition} if
$f_g(p_1,q, M,D)\ge f_g(p_2,q, M,D)$ for all $0\le p_1\le p_2\le 1$,
$D>0$, and $q\in [0,1]$.

We have the following algorithm for variant connection topologies
for ESs.

\vskip 10pt


\begin{algorithm}[t]
	\caption{ Randomized Algorithm} 
	\hspace*{0.02in} {\bf Input:}
	A parameter set $M$ (see definition 2, probability parameter $p$, maximum transmission range $m_0$, initial vehicle speed $v_0$, time threshold $t_0$, average number of vehicles density $b$, parameters $\gamma,\epsilon\in (0,1)$.\\
	\hspace*{0.02in} {\bf Output:} 
	$d$
	\begin{algorithmic}[1]
		
		\State Let $d_1=m_0$, $i=1$, $\lambda_0=0.1$, and $\delta=\gamma/3$; 
		\State Select the least integer $h$ such that $(1+\epsilon)^hm_0\ge 2D$;
		\State Select the least integer $t$ such that $he^{-{t\delta^2\over
				2}}\le \lambda_0$;
		\State \label{loop-start}
               Repeat
		\State \qquad Let $X_j$=R$_g(d_i,q, M, D)$ for $j=1,2,\cdots, t$;
		\State \qquad Compute $S=\sum_{j=1}^t X_j$;
		\State \qquad Let $d_{i+1}=d_i(1+\epsilon)$ and $i=i+1$;
		\State \label{test-line} 
	  	        Until $S<(p-\delta)t$ or $d_i>2D$;
		\State  $d=d_{i-1}$;

	\end{algorithmic}
\end{algorithm}

\vskip 10pt


{\bf Theorem 1.} Assume that $M$ is a parameter set,
and $g(.)$ indicates the ES connection topology. They satisfy the
monotonic condition. Let $D$ be the parameter that controls the
range for message transmission from the accident site. Let
parameters $p,q$ be in $[0,1]$, and $\gamma$ be in  $[0,p]$. Then
it exists a given probability meet the connectivity of IoVs under following condition. It gives a distance $d$ with
${f_g(p+\gamma,q,M,D)\over 1+\epsilon}\le d\le f_g(p-\gamma,q, M,D)$
in
 time
 $\complexity$
where
 $n_D$ is the number of vehicles on the road to the first message
 site of distance at most
$D$, and $T(M, n_D, h_D)$ is the time of generation and simulation
of a random event $R_g(.)$ for the system of parameters $M$, $n_D$
vehicles, and $h_D$ is the number of ESs to the accident
 site of distance at most
$D$. Furthermore, $T(M, n_D, h_D)$ is not decreasing for both $n_D$
and $h_D$.

We note that a concrete computational time complexity for $T(M, n_D,
h_D)=O(n^2\log n)$ with $n=n_D+h_D$ will be given at
section~\ref{simulation-algorithm-section}, where we develop a
simulation algorithm.

\begin{proof} Let parameters
$m_0$, $i$, $\lambda_0$,  $i$, $\delta=\gamma/3$, and $X_j$
 be defined as in Algorithm.1.


The number of cycles of the loop (lines \ref{loop-start} to 8 in the algorithm)  is bounded by $h$ with
$(1+\epsilon)^hm_0\ge 2D$. Thus,

\begin{eqnarray}
h=\ceiling{\ln (2D/m_0)\over \ln (1+\epsilon)} =O({1\over
\epsilon}{\ln{{D\over m_0}}}).\label{select-h-eqn-org}
\end{eqnarray}

Select parameter $t$ for the number of random events on a highway as
follows
\begin{eqnarray}
t&=&\ceiling{2\ln({h\over \lambda_0})\over \delta^2}\label{select-h-eqn}\\
&=&O({1\over \delta^2}(\ln\ln({D\over m_0})+\ln{1\over
\epsilon}))\label{select-h-eqn2}\\
 &=&O({1\over
\gamma^2}(\ln\ln({D\over m_0})+\ln{1\over
\epsilon}))\label{select-h-eqn3}.
\end{eqnarray}

By equation (\ref{select-h-eqn}), the selection of parameters $h$
and $t$ makes
\begin{eqnarray}
he^{-{1\over 2}t\delta^2}\le \lambda_0.\label{lambda-ineqn}
\end{eqnarray}

If $d_i<f_g(p+\gamma,M,D)$, then with probability at most
$e^{-{1\over 2}t\delta^2}$,  $\sum_{i=1}^tX_i<
(p+\gamma-\delta)t=(p+2\delta)t$  by
Proposition 2. If $\sum_{i=1}^tX_i\ge
(p+\gamma-\delta)n=(p+2\delta)t$, it fails the test of
line 8 in the algorithm and enters cycle $i+1$ for
testing $d_{i+1}$. Thus, with probability at most $he^{-{1\over
2}t\delta^2}$, we fail to have an output  $d\ge
{f_g(p+\gamma,M,D)\over 1+\epsilon}$.

If $d_i\ge f_g(p-\gamma, q, M,D)$ (note $f_g(p-\gamma,M,q, D)\ge
f_g(p+\gamma,M,D)$ by the monotonic condition of $M$), then we have
$\sum_{i=1}^tX_i> (p-\gamma+\delta)t=(p-2\delta)t$ with probability
at most $e^{-{1\over 2}t\delta^2}$ (by
Proposition 2). If $\sum_{i=1}^tX_i\le
(p-\gamma+\delta)t=(p-2\delta)t$, it passes the test at
line 8 of the algorithm, and returns $d=d_{i-1}$. If
$i$ is the least integer with $d_i\ge f_g(p-\gamma, q, M,D)$, then
$d_{i-1}\le f_g(p-\gamma, q, M,D)$.  Thus, with probability at most
$e^{-{1\over 2}t\delta^2}$, we fail to have an output $d\le
f_g(p-\gamma,M,D)$.

By inequality~\ref{lambda-ineqn}, with probability at most
$(h+1)e^{-{1\over 2}t\delta^2}\le 2he^{-{1\over 2}t\delta^2}\le
2\lambda_0$, we fail to output $d$ with ${f_g(p+\gamma,M,D)\over
1+\epsilon}\le d\le f(p-\gamma, M,D)$.

Each cycle samples sufficient $t$ random events. The total
number of cycles in the loop is at most $h$.  The maximal number of
ESs is at most ${2D\over m_0}$ as the distance of two consecutive
ESs should not be less than $m_0$. The total amount time is $t\cdot h\cdot T(M, n_D,
h_D)$, which
 matches the complexity claim in the theorem by equations (\ref{select-h-eqn-org}) and
(\ref{select-h-eqn})-(\ref{select-h-eqn3}).

\end{proof}

The monotonic condition is satisfied for both connected ESs and
unconnected ESs. The algorithm is applied for Connected ESs when
$R_{g_c}(d_i,M, D)$ is used in the simulation.

{\bf Corollary 1.} Assume that $M$ is a parameter set for highway traffic with connected ESs with connection topology $g_c(.)$. Let $D$ be the parameter that controls
the range for message transmission from the accident site. Let
parameters $p,q$ be in $[0,1]$, and $\gamma$ be in  $[0,p]$. Then
there is an approximation algorithm that gives a distance
$d$ and meets ${f_{g_c}(p+\gamma,M,D)\over 1+\epsilon}\le d\le
f_{g_c}(p-\gamma, M,D)$. The time is $\complexity$, where  $n_D$ is
the number of vehicles on the road of length $D$, and $T(M, n_D,
h_D)$ is the time of simulation for the system of parameters $M$,
$n_D$ vehicles, and $h_D$ is the number of ESs on a road of length
$D$.


The algorithm is applied for Connected ESs when $R_{g_u}(d_i,M, D)$
is used in the simulation.

{\bf Corollary 2.} Assume that $M$ is a parameter set for highway traffic with unconnected ESs
with connection topology $g_u(.)$. Let $D$ be the parameter that
controls the range for message transmission from the accident site.
Let parameters $p,q$ be in $[0,1]$, and $\gamma$ be in  $[0,p]$.
Then there is an approximation algorithm that gives a
distance $d$ and meets ${f_{g_u}(p+\gamma,M,D)\over 1+\epsilon}\le
d\le f_{g_u}(p-\gamma, M,D)$. The time is  $\complexity$, where
$n_D$ is the number of vehicles on the road of length $D$, and $T(M,
n_D, h_D)$ is the time of simulation for the system of parameters
$M$, $n_D$ vehicles, and $h_D$ the number of ESs on a road of
length $D$.

\section{An Algorithm for
Simulation}\label{simulation-algorithm-section}

In this section, we give an algorithm for simulation. It has a
rigorous analysis for both correctness and complexity. Our algorithm
can simulate a IoVs that has many vehicles with variant speeds, and
multiple lanes on the highway. It has a reasonable computational
complexity that makes it implementable by software .

We first give a brief description of the algorithm.  Each ES is
considered as a vehicle of speed zero. The algorithm is recursive
via linear order of the times for the vehicles receiving the
message. Two B-trees $T_T$ and $T_N$ hold the list of vehicles to
receive the message within time $t_0$. $T_T$ is used to hold the set
of vehicles by their time to receive the message, and $T_N$ is used
to hold the set of vehicles by their names.  Our algorithm
identifies the set of vehicles $P_i$ that can receive the message
from the vehicle $c_i$ after $c_i$ has got the message. A vehicle
$c_i$ in $T_T$ with least time $t_i$ is put into the output list
$L_2$. For each vehicle $c_i$, calculate the time $t_j$ to receive
the message directly from $c_i$ for each $c_j\in P_i$.  Delete $c_i$
from both $T_T$ and $T_N$.  If $T_T$ and $T_N$ already contain
$c_j\in P_i$, it will be replaced by the new time $t_j$ if it is
earlier than the old time to receive the message for $c_j$. The set
of vehicles in $P_i$ will be inserted into two B-trees $T_T$ (by the
order of $t_j$) and $T_N$ (by the order of their IDs). There is a
two directional link for the two nodes of each vehicle in $T_T$ and
$T_N$.


{\bf Definition 6.} Let $g(.)$ be a connection topology for ESs on a highway. A ES $x$
{\it directly connects} to another ES $y$ if they are connected
$g(x)=g(y)$, and there is no ES $z$ between $x$ and $y$ with
$g(x)=g(z)$.

By the definition of direct connection, one ES connects at most two
ESs on a highway.

\vskip 10pt

\begin{algorithm}[t]
	\caption{ Simulation Algorithm} 
	\hspace*{0.02in} {\bf Input:}
	parameter $t_0$ for the time delay, the positions of ESs,
	and vehicles with speed.\\
	\hspace*{0.02in} {\bf Output:} 
	the list $L$ of vehicles and ESs that receive the message
	within time $t_0$.
	\begin{algorithmic}[1]
		
		\State Let each ES is treated as a vehicle of speed zero. 
		\State For each car $c_i$, find the set of vehicles $P_i$ that can receive message from $c_i$ within time $t_0$.
		\State Identify the first vehicle $c_k$ to receive the message, put it into $T_N$ and $T_T$, and set up a link from $T_N$ to $T_T$ for this vehicle in both trees.
		\State Build a B-tree $T_N$ to save the cars by the linear order of their names.
		\State Build a B-tree $T_T$ to save the cars by the linear order of their time to receive message.
		\State Let $L_2$ be an empty list.
		\State Put the car in $L_1$ into $T_N$ and $T_T$, and set up a link from $T_N$ to $T_T$ for the same vehicle.
		\State Repeat 
		\State \qquad for each vehicle $c_i$ with least time to receive the message in $T_T$,
		\State \qquad \qquad delete $c_i$ from  $T_T$ and $T_N$, and put it into a list $L_2$.
		\State \qquad\qquad put all vehicles in $P_i$ into $T_N$ and $T_T$, set up a link from $T_N$ to $T_T$ for the same vehicle, and delete the existing vehicle  if its time to receive the message is later, and insert the new time.
		\State Until $T_T$ is empty.
		\State $L=L_2$.
	\end{algorithmic}
\end{algorithm}

\vskip 10pt

{\bf Theorem 2.} There is an $O(P(t_0)n\log n)$ time algorithm to determine the set
of vehicles that will receive message, where $P(t_0)$ is the largest
number of vehicles that one vehicle or ES can directly pass the
message to other vehicles or ESs on the road, and $n$ is the total
number of vehicles and ESs .

We only let at most two ESs directly receive message from one node.
They can continue pass the message to the others connected to them.
This controls the $P(t_0)$ to be small.

\begin{proof}
The correctness for this algorithm can be obtained by a simple
induction for the number of vehicles on the road. Each ES is
treated as a vehicle of speed zero in the algorithm. Each ES passes
the message directly to its neighbor ESs if they are connected,  or
those vehicles and ESs in the range radio transmission.
 It is trivial
when there is only one vehicle on the road. Assume that the
algorithm works for the case that there are $n$ vehicles such that
each vehicle is added to the list $L_2$ by the earliest time
receiving the message. Consider the case of $n+1$ vehicles. Let
$c_{n+1}$ be the rightmost vehicle on the road. We discuss the
following cases.

The vehicle $c_{n+1}$ is reachable by neither $ES$ nor
other vehicles. It follows from the inductive hypothesis.


Case 2. The vehicle $c_{n+1}$ is reachable first by another vehicle
$c_{i}$ at time $t_{n+1}$. It will be considered in $P_i$. When
$c_i$ is added to $L_2$, $c_{n+1}$ will be in $P_i$ and will be
added to the list $L_2$ according to time $t_{n+1}$. After vehicle
$c_i$ is added $L_2$, it will be added to neither $L_2$ nor B-tree.
It becomes the case of $n$ vehicles on the road. The other vehicles
with message passed from $c_{n+1}$ follows from the inductive
hypothesis.

Therefore, the algorithm works for the case with $n+1$ vehicles.
This proves the correctness of the algorithm.

Each vehicle can forward message to at most $P(t_0)$ vehicles. The
B-tree operation takes $O(\log n)$ time for inserting and deleting.
Each vehicle has at most $O(P(t_0)$ times to do B-tree operations.
Therefore, the total time is $O(P(t_0)n\log n)$.
\end{proof}

{\bf Corollary 3.} It exists an $O(n^2\log n)$ time algorithm to determine the set of
vehicles that will receive message, where $P(t_0)$ is the largest
number of vehicles that one vehicle or ES can directly pass the
message to other vehicles or ESs on the road, and $n$ represents the total
number of vehicles and ESs .

The generation of a random traffic takes $O(n)$ for a piece of
highway with $n$ vehicles and ESs according to a  system of
parameters $M$ for highway traffic.

{\bf Theorem 3.} Assume that $M$ is a parameter set,
and $g(.)$ is the ES connection topology. They satisfy the
monotonic condition. Let parameters $p,q$ be in $[0,1]$. Then there
is an approximation algorithm such that it gives  a distance $d$
with ${f_g(p+\gamma,q,M,D)\over 1+\epsilon}\le d\le f_g(p-\gamma,q,
M,D)$ in
 time
 $\complexitytwo$
where $D$ is the length to be considered for the
message transmission, $n_D$ is the number of vehicles on the road of
length $D$, and $T(M, n_D, h_D)$ is the time of simulation for the
system of parameters $M$, $n_D$ vehicles, and $h_D$ is the number of
ESs on a road of length $D$. Furthermore, $T(M, n_D, h_D)$ is not
decreasing for both $n_D$ and $h_D$.

\begin{proof}
It follows from Theorem 1 and
Corollary 3.
\end{proof}

\section{Simulation Results}\label{simlation-section}

\par There is no algorithm that can calculate the optimal solution during polynomial running time since the problem of ESs placement in IoVs is NP-hard \cite{li2015delay}. What we can do is approaching the approximation optimal solution as much as possible. It is unnecessary to cover all of the nodes to complete connectivity in practical application. We focus on the relation about the placement distance of ESs or the number of ESs according to the connection probability of vehicular network.

\par For each highway scenario, we can calculate the approximate optimal solution of ES deployment by this scheme. This experimental scenario is set as follows. According to the daily traffic volume of WUE highway in China, we calculate the average vehicle capacity of the highway. That is 1060. It means that there are 1060 vehicles on the highway. We consider two scenarios of vehicle density. When the vehicle node is 1060, it is a general scenario, and when the vehicle node is 530, it is a sparse scenario. Where vn is the number of vehicles and vES is the number of deployed ESs. The node communication adopts the DSRC. The maximum distance of node transmission is 200 meter. We take 200 meter as the common default value $m_0$

\par The ESs deployed has the same interval. The initial position of vehicle nodes is randomly on highway scenario. The simulation results show that the IoVs connectivity rate increases with the total number of ESs, as shown in Fig.1, Fig.2, Fig.4, and Fig.6.

Define the {\it direct connectivity probability} of vehicle
with ESs is the number of vehicles on the highway directly connected to ESs divided 
by the total number of vehicles on the highway.

Define the {\it indirect connectivity probability} of vehicle
with ESs is the number of vehicles on the highway that can communicate with ESs via the relay of some other vehicles divided 
by the total number of vehicles on the highway.

Define the {\it connectivity probability} of vehicle
with ESs is the number of vehicles on the highway is the sum of direct connectivity probability of indirect connectivity probability.

\par When $m_0$ is 200 meters, the direct connectivity probability of vehicle
to ESs increases almost linearly as the number of ESs gets larger 
The direct connectivity probability of vehicle
to ESs is much larger than the indirect connectivity probability of vehicle
to ESs.
On the other hand, the indirect connectivity probability of vehicle
with ESs is not linearly increasing with the increasing number of ESs.

For vn=530, indirect connectivity probability of vehicle
with ESs becomes maximum when ESs =650.

The indirect connectivity probability decreases when the number of ESs
is increased. The reason is that the number of vehicles directly
connected with ESs increases when the number of ESs is increased. 
The connectivity probability goes up slowly.

When the number of deployed ESs reaches
$650$, the connectivity probability is 0.775, as shown in
Fig.1. We can consider vES=650 as an approximation for the optimal solution in the
case.
For vn=1060,  vES=600 is an approximation for the optimal solution in the
case, which is  shown in Fig2.

When the number of vehicles is fixed, the
direct connectivity probability of vehicle with ESs is almost
constant regardless of the number of vehicles, which is shown in Fig.3, Fig.5, and Fig.7.
However, the indirect connectivity probability of vehicle with ESs
almost linearly increases with the increasing number of ESs.

\begin{figure}[htp]
\centering
\includegraphics[width=3.0in,height=3.0in,clip,keepaspectratio]{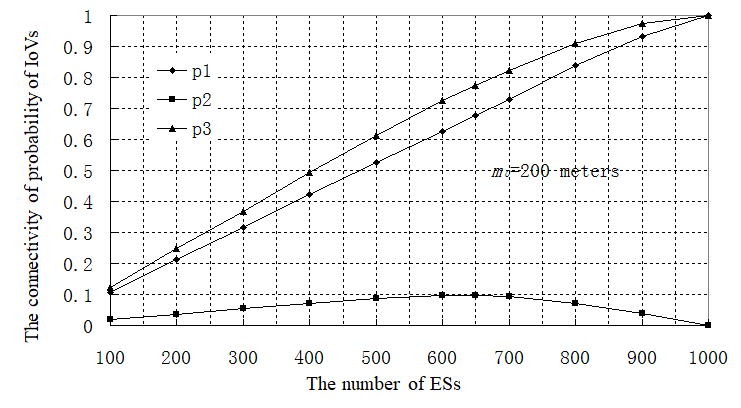}
\caption{ The connectivity probability of IoVs for the number of
ESs with vn=530.\protect \\ p1:The probability of vehicles directly
connected ES.\protect \\  p2: The probability of vehicles
indirectly connected ES.\protect \\ p3: The total probability of
vehicle connected ES.}
\end{figure}


\begin{figure}[htp]
\centering
\includegraphics[width=3.0in,height=3.0in,clip,keepaspectratio]{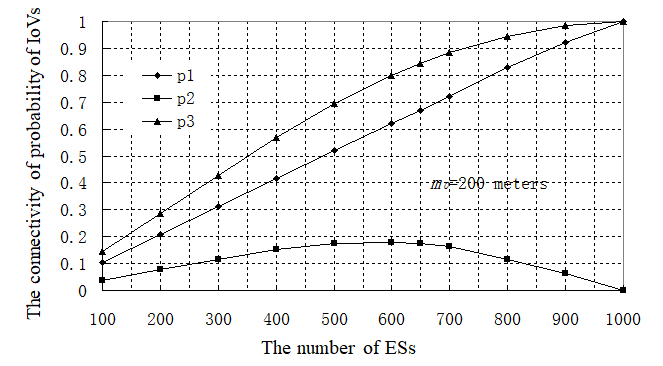}
\caption{\small The connectivity probability of IoVs for
the number of ESs with vn=1060.\protect \\ p1:The probability of
vehicles directly connected ES.\protect \\  p2: The probability of
vehicles indirectly connected ES.\protect \\ p3: The total
probability of vehicle connected ES.}
\end{figure}

\begin{figure}[htp]
\centering
\includegraphics[width=3.0in,height=3.0in,clip,keepaspectratio]{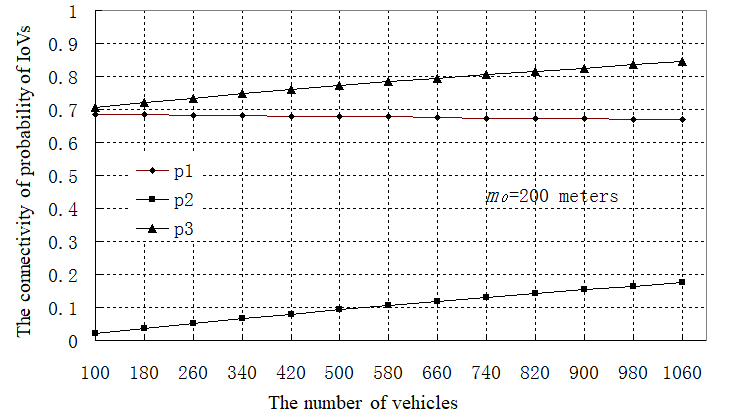}
\caption{ The connectivity probability of IoVs for the number of
vehicles with vES=650.\protect \\ p1:The probability of vehicles
directly connected ES.\protect \\  p2: The probability of vehicles
indirectly connected ES.\protect \\ p3: The total probability of
vehicle connected ES.}
\end{figure}

\par When vn =1060, the connected probability has the similar trends with vn=530.
But, the number of ESs need to deploy is significant reduction. The
approximation optimal solution is vES=50 and vES=90 with the
vn=1060 and vn=530 respectively. The connectivity probability is up
to 0.806  with vES=50, vn = 530 according to the $m_0$ is 200m, vES=
680, vn=530.

\begin{figure}[htp]
\centering
\includegraphics[width=3.0in,height=3.0in,clip,keepaspectratio]{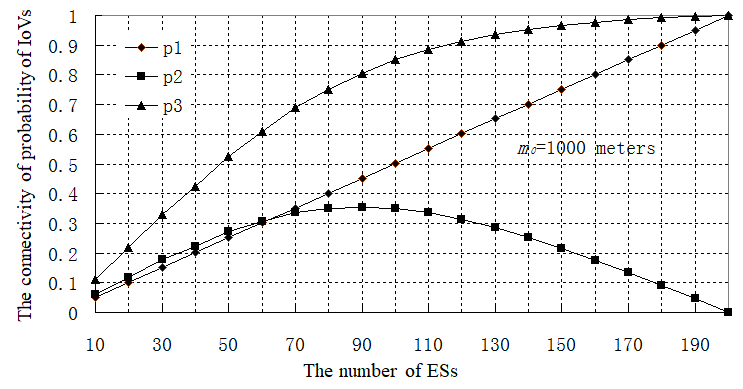}
\caption{ The connectivity probability of IoVs for the number of
ESs with vn=530.\protect \\ p1:The probability of vehicles directly
connected ES.\protect \\  p2: The probability of vehicles
indirectly connected ES.\protect \\ p3: The total probability of
vehicle connected ES.}
\end{figure}

\begin{figure}[htp]
\centering
\includegraphics[width=3.0in,height=3.0in,clip,keepaspectratio]{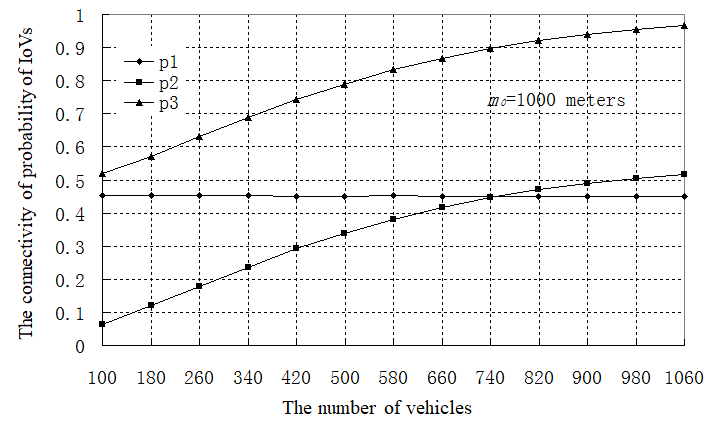}
\caption{ The connectivity probability of IoVs for the number of
vehicle with vES=90.\protect \\ p1:The probability of vehicles
directly connected ES.\protect \\ p2: The probability of vehicles
indirectly connected ES.\protect \\ p3: The total probability of
vehicle connected ES.}
\end{figure}

\begin{figure}[htp]
\centering
\includegraphics[width=3.0in,height=3.0in,clip,keepaspectratio]{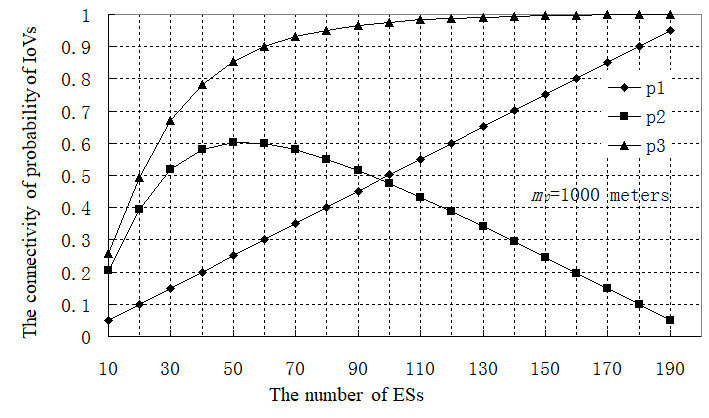}
\caption{ The connectivity probability of IoVs for the number of
ESs with vn=1060.\protect \\ p1:The probability of vehicles
directly connected ES.\protect \\  p2: The probability of vehicles
indirectly connected ES.\protect \\ p3: The total probability of
vehicle connected ES.}
\end{figure}

\begin{figure}[htp]
\centering
\includegraphics[width=3.0in,height=3.0in,clip,keepaspectratio]{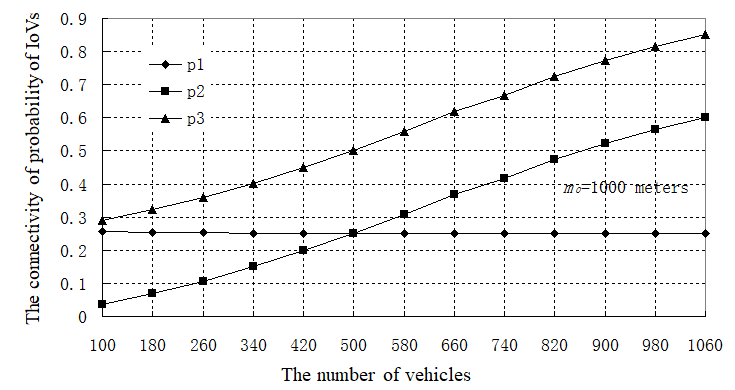}
\caption{ The connectivity probability of IoVs for the number of
vehicle with vES=50.\protect \\ p1:The probability of vehicles
directly connected ES.\protect \\  p2: The probability of vehicles
indirectly connected ES.\protect \\ p3: The total probability of
vehicle connected ES.}
\end{figure}

\par Fig.8 and Fig.9 show that the connectivity probability of IoVs for the number of ESs with vn=1060, vspeed =108 km/h, and vspeed = 216 km/h: 1) The transmission distance of vehicles is 200 meters; 2) The transmission distance of vehicles is 1000 meters. We can see that the speed of vehicles have little impact on the connectivity probability.

\begin{figure}[htp]
\centering
\includegraphics[width=3.0in,height=3.0in,clip,keepaspectratio]{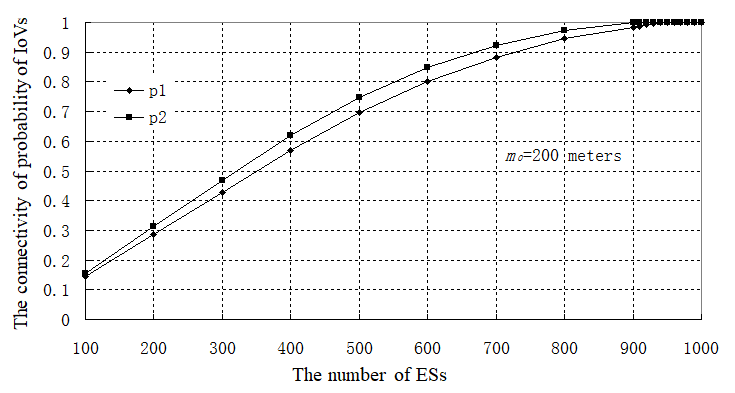}
\caption{ The connectivity probability of IoVs for the number of ESs with $m_0$=200 meters, vn=1060 , vspeed = 108 km/h, and vspeed=216km/h.\protect \\ p1:The connectivity probability of vehicles with vspeed = 108 km/h.\protect \\  p2: The connectivity probability of vehicles with vspeed = 216 km/h.}
\end{figure}

\begin{figure}[htp]
\centering
\includegraphics[width=3.0in,height=3.0in,clip,keepaspectratio]{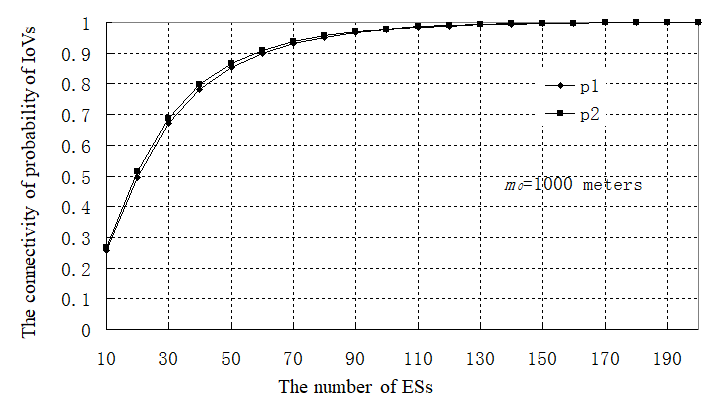}
\caption{ The connectivity probability of IoVs for the number of ESs with $m_0$=1000 meters, vn=1060 , vspeed = 108 km/h, and vspeed=216km/h.\protect \\ p1:The connectivity probability of vehicles with vspeed = 108 km/h.\protect \\  p2: The connectivity probability of vehicles with vspeed = 216 km/h.}
\end{figure}

\par We compared the proposed scheme with ODEL\cite{mehar2015optimized}. As shown in Fig.10 and Fig.11, we find that the deployment cost of ODEL scheme is higher than that of the proposed scheme with the same connectivity probability of IoVs. It is because ODEL method needs to deploy more ESs to satisfy the requirements to reduce the routing delay caused by dynamic network topology in IoVs. The scheme proposed in this paper focuses on the fact that the deployed edge servers can cover more segments in road, so the deployment cost can be reduced.

\begin{figure}[htp]
	\centering
	\includegraphics[width=3.0in,height=3.0in,clip,keepaspectratio]{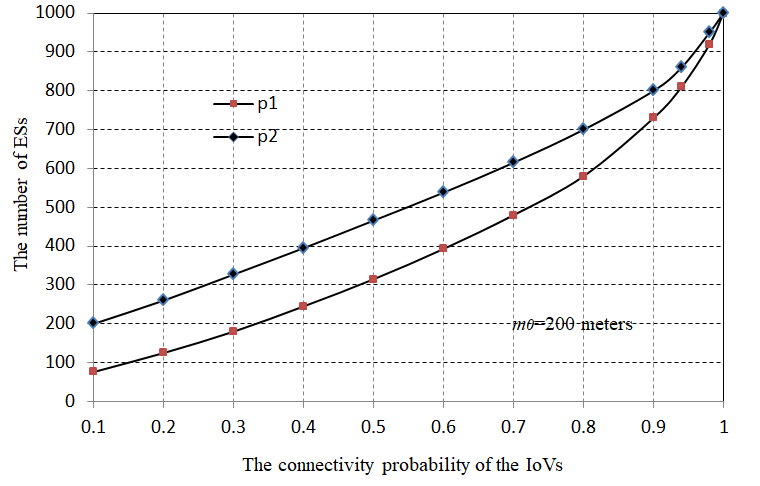}
	\caption{ The number of ESs vs The connectivity probability of IoVs with $m_0$=200 meters.\protect \\ p1: The number of ESs in the proposed scheme.\protect \\  p2: The number of ESs in ODEL.}
\end{figure}

\begin{figure}[htp]
	\centering
	\includegraphics[width=3.0in,height=3.0in,clip,keepaspectratio]{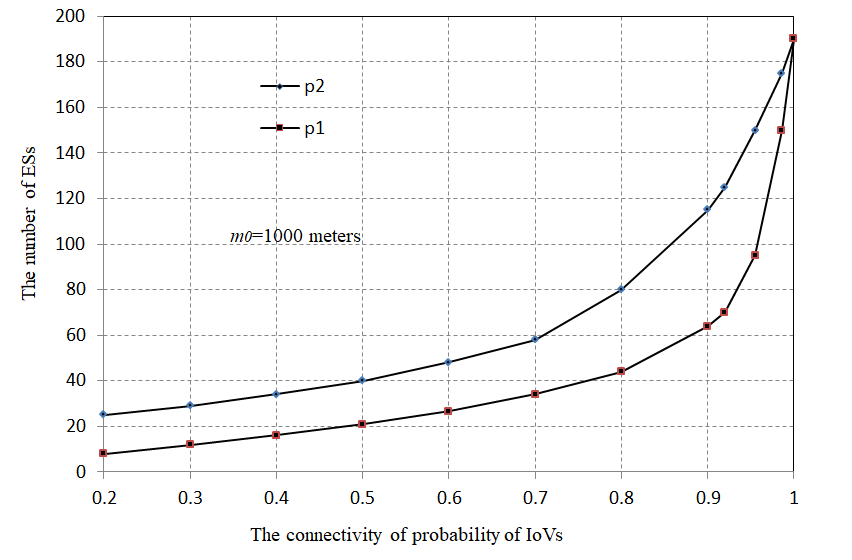}
	\caption{ The number of ESs vs The connectivity probability of IoVs with $m_0$=1000 meters.\protect \\ p1:The number of ESs in the proposed scheme.\protect \\  p2: The number of ESs in ODEL.}
\end{figure}

\section{Conclusion}
\par In this paper, we investigated the scheme of edge server deployment in IoVs, which enables edge computing to be implemented in IoVs for the application of blockchain. In the scheme, the roadside units are considered as edge servers of edge computing. We introduce a randomized method to develop an approximation algorithm for edge server deployment. Our goal is to deploy a minimal number of edge servers while vehicle nodes can be linked to at least one of the ESs. It obtains an approximation for the optimal deployment distance to ensure the message can be transmitted to ESs from the source site via the IoVs. Moreover, we design an efficient algorithm to simulate IoVs environment with vigorous theoretical proof for its correctness and complexity. The simulation results show the number of ESs depends on some parameters such as wireless transmission distance, the density of vehicles, etc. Finally, we compared the proposed scheme with other schemes in terms of the deployment cost for the connectivity probability of IoVs.

\section*{Acknowledgment}

\par This work is supported by the National Science Foundation of China(No.61662039), Science and technology project of Jiangxi Provincial Department of Education (No. GJJ170967), Jiangxi Key Natural Science Foundation (No. 20192ACBL20031), Project of Teaching Reform in Jiujiang University(No. XJJGYB-19-47).

\ifCLASSOPTIONcaptionsoff
  \newpage
\fi



%

 \bibliographystyle{unsrt}
\bibliography{mybibliography}

\end{document}

%% file: mac90.tex
\newcounter{savenumi}

\newtheorem{theoremfoo}{Theorem}

\newtheorem{propositionfoo}[theoremfoo]{Proposition}

\newtheorem{lemmafoo}[theoremfoo]{Lemma}

\newtheorem{conjecturefoo}[theoremfoo]{Conjecture}

\newtheorem{corollaryfoo}[theoremfoo]{Corollary}

\newtheorem{exercisefoo}{Exercise}

\newtheorem{openfoo}[theoremfoo]{Question}

\newtheorem{nttn}[theoremfoo]{Notation}

\newtheorem{dfntn}[theoremfoo]{Definition}

\newenvironment{proof}
    {\pagebreak[1]{\narrower\noindent {\bf Proof:\quad\nopagebreak}}}{\QED}

\renewcommand{\theenumi}{\roman{enumi}}

\newcommand{\ceiling}[1]{\left\lceil#1\right\rceil}

\renewcommand{\tt}{{\rm tt}}    
\def\nre.{$n$\/-r.e.}

\newtheorem{factfoo}[theoremfoo]{Fact}

\newcommand{\squeeze}{
\textwidth 6in
\textheight 8.8in
\oddsidemargin 0.2in
\topmargin -0.4in
}

\newtheorem{propertyfoo}[theoremfoo]{Property}

\makeatletter

\def\@makechapterhead#1{ \vspace*{50pt} { \parindent 0pt \raggedright 
 \ifnum \c@secnumdepth >\m@ne \huge\bf \@chapapp{} \thechapter. \par 
 \vskip 20pt \fi \Huge \bf #1\par 
 \nobreak \vskip 40pt } }

\def\@sect#1#2#3#4#5#6[#7]#8{\ifnum #2>\c@secnumdepth
     \def\@svsec{}\else 
     \refstepcounter{#1}\edef\@svsec{\csname the#1\endcsname.\hskip 1em }\fi
     \@tempskipa #5\relax
      \ifdim \@tempskipa>\z@ 
        \begingroup #6\relax
          \@hangfrom{\hskip #3\relax\@svsec}{\interlinepenalty \@M #8\par}
        \endgroup
       \csname #1mark\endcsname{#7}\addcontentsline
         {toc}{#1}{\ifnum #2>\c@secnumdepth \else
                      \protect\numberline{\csname the#1\endcsname}\fi
                    #7}\else
        \def\@svsechd{#6\hskip #3\@svsec #8\csname #1mark\endcsname
                      {#7}\addcontentsline
                           {toc}{#1}{\ifnum #2>\c@secnumdepth \else
                             \protect\numberline{\csname the#1\endcsname}\fi
                       #7}}\fi
     \@xsect{#5}}

\def\@begintheorem#1#2{\it \trivlist \item[\hskip \labelsep{\bf #1\ #2.}]}

\def\@opargbegintheorem#1#2#3{\it \trivlist
      \item[\hskip \labelsep{\bf #1\ #2\ (#3).}]}

\makeatother

\newif\ifshortconferences
\shortconferencesfalse
\newif\ifmediumconferences
\mediumconferencesfalse

\def\ending#1{{\count1=#1\relax
\count2=\count1
\divide\count2 by 100
\multiply\count2 by 100
\advance\count1 by -\count2
\ifnum\count1=11
th%
\else \ifnum\count1=12
th%
\else \ifnum\count1=13
th%
\else 
\count2=\count1
\divide\count1 by 10
\multiply\count1 by 10
\advance\count2 by -\count1
\ifnum\count2=1
st%
\else \ifnum\count2=2
nd%
\else \ifnum\count2=3
rd%
\else th%
\fi\fi\fi\fi\fi\fi
}}

\def\Proceedingsofthe{\ifshortconferences Proc.\else\ifmediumconferences Proc.\else Proceedings of the\fi\fi}

\newcounter{confnum}

\def\conf#1#2{%
\setcounter{confnum}{#2}%
\addtocounter{confnum}{-\csname #1zero\endcsname}%
\ifnum\value{confnum}=1%
\expandafter\ifx\csname #1One\endcsname\relax%
\Proceedingsofthe\ \arabic{confnum}\ending{\value{confnum}}\ \csname #1name\endcsname%
\else \csname #1One\endcsname\fi%
\else%
\Proceedingsofthe\
\arabic{confnum}\ending{\value{confnum}}\ \csname #1name\endcsname\fi}

\def\qsym{\vrule width0.7ex height0.9em depth0ex}

\newif\ifqed\qedtrue

\def\noqed{\global\qedfalse}

\def\qed{\ifqed{\penalty1000\unskip\nobreak\hfil\penalty50
\hskip2em\hbox{}\nobreak\hfil\qsym
\parfillskip=0pt \finalhyphendemerits=0\par\medskip}\fi\global\qedtrue}

\makeatletter
\def\eqnqed{\noqed
	\def\@tempa{equation}
	\ifx\@tempa\@currenvir\def\@eqnnum{\qsym}%
	\addtocounter{equation}{-1}\else%
    \def\@@eqncr{\let\@tempa\relax
    \ifcase\@eqcnt \def\@tempa{& & &}\or \def\@tempa{& &}%
      \else \def\@tempa{&}\fi
     \@tempa {\def\@eqnnum{{\qsym}}\@eqnnum}
     \global\@eqnswtrue\global\@eqcnt\z@\cr}\fi}

\def\eqnlabel#1#2{\if@filesw {\let\thepage\relax%
   \def\protect{\noexpand\noexpand\noexpand}%
   \edef\@tempa{\write\@auxout{\string
      \newlabel{#2}{{{#1}}{\thepage}}}}%
   \expandafter}\@tempa%
   \if@nobreak \ifvmode\nobreak\fi\fi\fi%
	\def\@tempa{equation}
	\ifx\@tempa\@currenvir\def\theequation{{#1}}%
	\addtocounter{equation}{-1}\else%
    \def\@@eqncr{\let\@tempa\relax
    \ifcase\@eqcnt \def\@tempa{& & &}\or \def\@tempa{& &}%
      \else \def\@tempa{&}\fi
     \@tempa {\def\@eqnnum{{#1}}\@eqnnum}
     \global\@eqnswtrue\global\@eqcnt\z@\cr}\fi}

\makeatother

\def\QED{\qed}

\makeatother

